# Data-driven risk analysis of unmanned aircraft system operations considering spatiotemporal characteristics of population distribution

**Soohwan Oh**
Department of Civil and Environmental Engineering, Korea Advanced Institute of Science and Technology, Daejeon, Republic of Korea, 34141
Email: suhwan@kaist.ac.kr

**Yoonjin Yoon, PhD, Corresponding Author**
Department of Civil and Environmental Engineering, Korea Advanced Institute of Science and Technology, Daejeon, Republic of Korea, 34141
Email: yoonjin@kaist.ac.kr

## ABSTRACT

One of the challenges of Unmanned Aircraft System (UAS) operations is to operate an unmanned aircraft with minimal risk to people on the ground. The purpose of this study is to define and measure such risks as population risk, by incorporating spatiotemporal changes in population density. Unlike previous studies, we use high-resolution de facto population data instead of residential population data to reflect the spatiotemporal characteristics of population distribution. Furthermore, we analyze the impact of mitigation measures based on population risk in the context of airspace management. We set a restricted airspace by using population risk and an acceptable level of safety. Scenario analysis of the study area in Seoul, South Korea provides a richer set of findings regarding spatiotemporal differences in restricted airspace. During the daytime, there are many restricted airspaces around commercial areas, but few around residential areas. Additionally, we observe the difference between restricting airspace based on population risk derived from the residential population and from the de facto population. These findings confirm the importance of accurately considering population density when assessing and mitigating the population risk associated with UAS operations. Sensitivity analysis also reveals the need to precisely estimate population density when estimating population risk with combinations of multiple parameter values. The proposed approach captures spatiotemporal characteristics of population distribution when assessing the population risk associated with UAS.

# 1. Introduction

The demand for Unmanned Aircraft Systems (UAS) will grow with various applications such as medical aid, food delivery, and package delivery (Doole et al., 2020; Kellermann et al. 2020; Rajendran et al. 2021). The expected growth in UAS demand has raised safety concerns associated with its use (Washington et al., 2017). One of the most serious concerns is the risk of collision between an Unmanned Aircraft (UA) and people on the ground (Clothier et al., 2018). To address the potential risk associated with UA flying over people, one simple but effective measure is to issue a temporary flight restriction (TFR) over a high-risk area, making the airspace inoperable for a certain period (Oh et al., 2020).

Because such TFR renders portions of airspace inoperable, the designation of airspace as TFR could have a significant impact on airspace operability (Cho and Yoon, 2018; Vascik et al., 2020). Therefore, it is of utmost importance to determine the reasonable extent and duration of TFR area based on quantitative metrics such as casualty risk associated with UAS operations. A number of studies have attempted to measure the casualty risk, but have primarily focused on the dynamics of individual UA, such as descent trajectory and impact area (Clothier et al., 2018; Primatesta et al., 2020). They had limited consideration of the environments exposed to UAS operations, including people exposed to UAS operations. For example, the aforementioned studies calculated the casualty risk based on residential population density. The limitation of this approach is that it does not consider people who are actually at risk of UAS operations. If TFR is designated solely based on residential population, areas with high residential population but low daytime population can be an example of inefficient restriction of airspace, and commercial districts with high daytime population but low residential population may not provide sufficient risk mitigation as intended. It is also important to note that, even though the casualty risk of individual UA flying on a specific route was calculated, the analysis about implementing mitigation measures based on the calculated risk was insufficient.

In this study, we define the casualty risk associated with UAS operations to people on the ground as $population\ risk$, incorporating spatiotemporal changes in population density. Using high-resolution de-facto population data, we measured $population\ risk$ for each time of day. In doing so, we calculated the casualty risk that reflects the spatiotemporal characteristics of population distribution rather than those derived from the residential population data. We then analyzed the impact of $population\ risk$ on airspace operability for airspace analysis. We define the airspace in which $population\ risk$ exceeds a certain threshold as the restricted airspace. We examined changes in the volume and shape of the restricted airspace during the day, considering the spatiotemporal characteristics of population distribution. We also investigated the effect of population density on the risk values for combinations of multiple parameter values through sensitivity analysis.

The remainder of this paper is organized as follows. In Section 2, the relevant literature is summarized. Section 3 describes risk assessment and the data used in this study. In Section 4, scenario analysis results for actual urban environments are presented and discussed. Section 5 presents the results of the sensitivity analysis. Finally, conclusions are presented in Section 6.

## 2. Related studies

The casualty risk of UAS operations has been assessed based on probabilistic risk assessment, which is a process of assessing a system's risk and reliability to improve its safety. Generally, it is used when there is insufficient data for an event with a low probability and high consequence. The casualty risk to people on the ground was evaluated in terms of the probability of a casualty occurring per flight hour (Blom et al., 2021; Poissant et al., 2020; Primatesta et al., 2020; Washington et al., 2017). This probability comprises the likelihood of an event and its consequences (Washington et al., 2017). The likelihood can be evaluated by considering the probability of an accident, the probability density to hit a location, the size of the impact area, and the probability of fatality. The consequence can be evaluated by considering sheltering effect and population density. A detailed description of how each element is considered in the relevant literature is as follows.

### 2.1 Probability of accident

There are several possible causes of a UA falling to the ground, including malfunctions and collisions with other aircrafts. However, estimating the probability of an accident by considering numerous possible causes is difficult. A major reason for this difficulty is that there are few real cases of UAS operations. Consequently, most studies have assumed a malfunctioning UA, irrespective of the event type (Ford et al., 2010; Petritoli et al., 2018; Stevenson et al., 2015).

A precise estimate of the probability of an accident caused by a malfunction may have been provided by the UA manufacturer, but not to the public. Because of this situation, relevant studies have assumed the probability based on expert opinions. Ford and McEntee (2010) assumed that the probability of a catastrophic accident is $10^{-5}$ per flight hour, and the probability of a hazardous accident is $10^{-4}$ per flight hour. A catastrophic accident is an event in which the UA cannot be controlled, whereas a dangerous accident can be controlled to a certain extent. Stevenson et al. (2015) assumed different probabilities for accidents based on the areas where the UA flies. The probability was assumed $10^{-5}$ per flight hour in suburban areas and $10^{-4}$ per flight hour in urban areas. Unlike the aforementioned studies, Petritoli et al. (2018) estimated probability based on actual data, rather than expert opinions.

### 2.2 Probability density to hit a location

Numerous researchers have studied the probability density of a UA descent trajectory hitting a specific location (Burke et al., 2011; Guglieri et al., 2014; La Cour-Harbo, 2019; La Cour-Harbo, 2020; Lum and Waggoner, 2011). Several factors influence the trajectory, including the initial state at the time of failure (e.g., position, velocity, and altitude), type of UA (e.g., fixed-wing, multi-rotor), mode of descent path (e.g., ballistic descent, uncontrolled glide descent, parachute descent), and environmental factors (e.g., wind).

Lum and Waggoner (2011) estimated the impact location based on the failure location and glide angle of the UA. The glide angle is the angle between the flight path and the horizontal plane as a flight descends to land. Because the glide angle affects the drag force of the UA, they assumed that it would glide to the ground with a maximum lift-to-drag ratio. Burke et al. (2011) assumed that the UA descended vertically such that the impact location was at the same location as the UA failed. Guglieri et al. (2014) adopted an uncontrolled glide descent model. The glide angle was set to 45° for the fixed-wing UA and 90° for the multirotor UA. La Cour-Harbo (2019) adopted a parachute descent model that assumes a vertical drop. Under the influence of wind, UA that has parachutes falls to the ground more slowly than UA that does not have one. La Cour-Harbo (2020) further considered that UA descended into a ballistic trajectory in a follow-up study. During ballistic descent, gravity and drag influence trajectory.

## 2.3 Size of impact area

Several factors determine the impact area, including the UA size and weight. Weibel and Hansman (2006) set the size of the impact area using the planform area of the UA. The area was set to $0.26 \text{ft}^2$ for micro UA weighing 0.14 lb, and $14 \text{ft}^2$ for mini UA weighing 9.6 lb. Clothier et al. (2007) estimated the size of the impact area based on the gliding area. The gliding area was determined using the equation of $2(r_p + r_d)\text{h}_p/\tan(\theta) + \pi(r_p + r_d)^2$, where $r_p$ and $\text{h}_p$ are the average radius and height of a person, respectively, $r_d$ is the radius of the vehicle, and $\theta$ is the impact angle on the ground. Burke et al. (2011) set the size of the impact area using the weight and wingspan of UA. Melnyk et al. (2014) estimated the size of the impact area based on the UA weight. They adopted the equation of $-2475.666\text{ft}^2 + 1.001\text{w}$, where w is the weight of the UA in pounds. The size of the impact area depends on the type, size, and weight of UA.

## 2.4 Probability of fatality

When a UA collides with a person on the ground, the UA transfers kinetic energy (KE) to the person. Because KE is a function of many factors (e.g. materials of the UA, initial altitude, and angle of impact), it is difficult to estimate. Accordingly, several studies selected KE based on a simple assumption (Burke et al., 2011; Clothier et al., 2007; Lum and Waggoner, 2011; Melnyk et al., 2014). Based on expert interviews, Clothier et al. (2007) assumed that the probability of a fatality was 100%. Melnyk et al. (2014) and Burke et al. (2011) assumed that the fatality rate was 100% when the KE exceeded 80 J and 92 J, respectively, based on historical accident and literature data. Lum et al. (2011) adopted the probability of fatality based on the type of UA. The probability of fatality was set to 100% for Predator B weighing more than 300 kg, and 50% for ScanEagle weighing 20 kg. In contrast to the aforementioned studies, Koh et al. (2018) calculated the impact energy on a dummy head when a UA was dropped from various heights using extensive simulations and experiments. Impact energy was then converted into the Abbreviated Injury Scale (AIS) to classify the injuries.

## 2.5 Sheltering effect

The sheltering effect determines the extent to which people in a given area are protected from collisions by the presence of trees or buildings. The sheltering effect has been considered in several studies (Blom et al., 2021; Melnyk et al., 2014; Stevenson et al., 2015; Weibel and Hansman, 2006). Weibel and Hansman (2006) assumed that the sheltering effect varies with the UA weight. The sheltering effect was set to 5% for micro UA weighing 0.14 pounds and 10% for mini UA weighing 9.6 pounds. Melnyk et al. (2014) estimated the sheltering effect from not only the KE but also roof material absorption data. Residential buildings are considered capable of protecting persons when the KE values are less than 2,700 J, and commercial buildings are considered capable of protecting persons when the KE values are less than 13,500J. Stevenson et al. (2015) assumed that the sheltering effect was dependent on the terrain. They selected a value of 75% for urban areas and 25% for wilderness terrain. Blom et al. (2021) assumed that 10% of the people are unprotected in urban areas. However, few studies have reflected the complex distribution of buildings in urban environments using high-resolution building data.

## 2.6 Population density

Most studies used census data to determine population density (Burke et al., 2011; Clothier et al., 2007; Ford and McEntee, 2010; Lum and Waggoner, 2011). Clothier et al. (2007) used census data to calculate population density; however, the spatial resolution was low because of the size of the census area. Using census data, Burke et al. (2011) accounted for the population density. Assuming the same population density values for each category, they classified the districts as unpopulated, sparsely populated, densely populated, or open-air assemblies. Ford and McEntee (2010) calculated population density based on building density assuming that building density correlated with population density. Lum and Waggoner (2011) assumed that the population density varies with regional characteristics. They set the value of population density as 5 per building in town areas, 20 per square kilometer in the field, and 10 per square kilometer in the forest. Based on this assumption, census data and satellite imagery were used to estimate the population density. However, there are limitations associated with the use of census data because they have a low spatial resolution and limited temporal information.

In summary, as presented in Table 1, relevant studies have assessed the risk that UAS operations pose to people by utilizing various impact area calculation methods and a fixed set of flight failure, fatality probability, and sheltering effect probabilities. In terms of population density, the number of people who at risk of UAS operations was derived from the residential population density. However, the residential population may not adequately account for people who are actually at risk of UAS operations during the daytime. There may be a minimum residential population in business districts while the number of people actually exposed to UAS operations can be maximized during the daytime. Therefore, the spatiotemporal characteristics of population distribution cannot be fully explained. Considering that, populated regions can appear from time to time and place to place, it is necessary to utilize de facto population data to the fullest extent and establish a reasonable flight restriction boundary that minimizes UAS operational risks. To this end, this study attempts to estimate reasonable casualty risk values for UAS operations by considering spatiotemporal characteristics of population distribution.

**Table 1.** Summary of relevant studies on the risk of UAS operations to people on the ground

| Literature | Probability of accident | Probability density to hit a location | Size of impact area | Probability of fatality | Sheltering effect | Population density |
|---|---|---|---|---|---|---|
| Weibel and Hansman (2006) | constant failure rate | vertical descent | planform area | KE-based probability | type-based probability | residential population |
| Clothier et al. (2007) | constant failure rate | - | gliding area | 100% | - | residential population |
| Burke et al. (2011) | constant failure rate | - | swept area | KE-based probability | KE-based probability | residential population |
| Lum and Waggoner (2011) | constant failure rate | uncontrolled glide descent | gliding area | mission-based probability | - | estimated population |
| Melnyk et al. (2014) | constant failure rate | - | weight-based equation | KE-based probability | material-based probability | estimated population |
| Primatesta et al. (2020) | constant failure rate | uncontrolled glide descent | gliding area | KE-based probability | terrain-based probability | residential population |
| Blom et al. (2021) | constant failure rate | ballistic descent | gliding area | KE-based probability | location-based probability | residential population |

## 3. Methodology

The goal of risk assessment is to minimize the expected operational risk by implementing mitigation measures (Melnyk et al., 2014; Poissant et al., 2020; Stevenson et al., 2015). One mitigation measure to address the potential risks associated with UA flying over people is to issue temporary flight restrictions over high-risk areas, making the airspace inoperable for a certain period (Oh et al., 2020). In the context of UAS traffic management, such restricted airspace can be considered a dynamic obstacle, and the formation of such obstacles can adversely impact airspace operability (Cho and Yoon, 2018; Vascik et al., 2020).

In this study, we define and measure the risk associated with UAS operations to people on the ground as $population\ risk$ to establish a reasonable containment boundary for a restricted airspace. We then set a restricted airspace based on $population\ risk$ with an acceptable level of safety, and the airspace was considered to be blocked by a dynamic obstacle. A dynamic obstacle represents the volume of stationary containment that is blocked over critical areas for a specific period of time. For example, the airspace above stadiums and densely populated areas may be temporarily restricted owing to $population\ risk$ and, therefore, may be regarded as dynamic obstacles in the air. We use a hexagonal cell as the spatial unit of airspace, which enables us UA to flight without acute and right angle turns (Yousefi and Donohue, 2004).

### 3.1 Data description

This study utilized de facto population data obtained from the Seoul Metropolitan Government Big Data Campus (Seoul Open Data Plaza, 2020). The data represent the number of people who are physically present in each neighborhood district of Seoul during each 1-hour interval. In contrast to residential population data, de facto population data can account for the daytime populations, and therefore can be used to analyze the population density during each 1-hour interval. Fig. 1 shows the difference between population density derived from residential and de facto population. We also used the height and shape information of buildings collected from the Ministry of Land, Infrastructure and Transport (MOLIT) of the Republic of Korea (National Spatial Data Infrastructure Portal, 2020).

(a) Residential population density      (b) De facto population density (13:00-14:00)

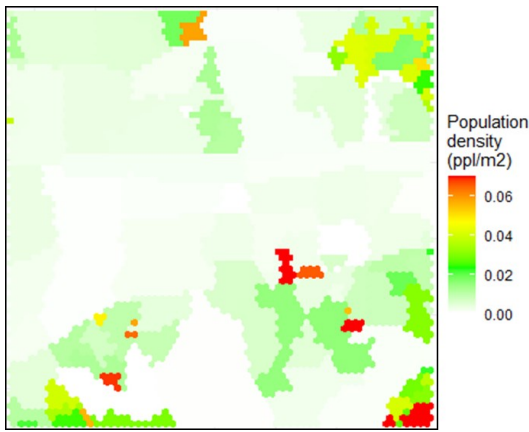 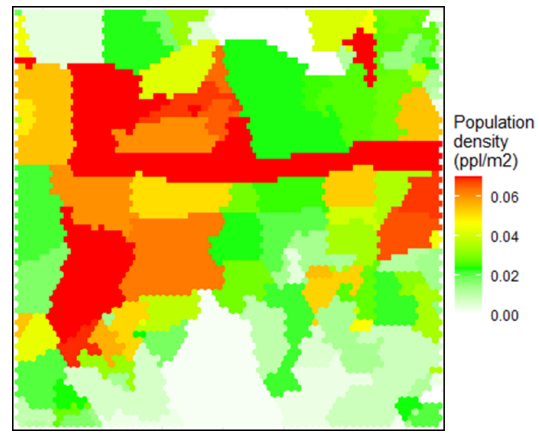

**Fig. 1.** Distribution of population density in the Jung district. (**a**) Residential population density; (**b**) De facto population density (13:00-14:00)

## 3.2 Generic metric of population risk

As mentioned in section 2, the risk of UAS operations to people on the ground is estimated in terms of the probability of causing a casualty per flight hour, which consists of the likelihood of an event and its consequences.

In terms of the likelihood, the probability of an accident, the probability density to hit a location, the size of impact area, and the probability of fatality can be considered. The probability of an accident can be assumed to be a constant failure rate, and the probability of fatality can be calculated using the KE. The probability density to hit a location is dependent on the direction of UA, but the direction of UA in the specified area is unknown. Primatesta et al. (2020) assumed that the direction of UA was equally distributed. Using this approach, the impact location is uniformly distributed. The size of the impact area depends on the size of the UA; however, a small UA is very small. Given the reasonable minimum spatial unit of airspace, we believe that the effect of the size of impact area is negligible.

With regards to the consequence, the sheltering effect and population density can be considered. Contrary to previous studies that consider the sheltering effect with a rather simplistic approach, we calculated the percentage of an area covered by buildings at a specific location using high-resolution building data. To fully investigate the spatiotemporal characteristics of population distribution, we extended the population perspective by adding temporal parameters derived from the de facto population data.

Considering this approach, we propose a generic metric of $population\ risk$ to measure the risk of UAS operations posed to people on the ground. $Population\ risk\ r_i^t(d,\ h)$ is defined as "the probability to cause a casualty per flight hour at location $i$ and time $t$ with respect to a crash of UA weighing $d$ falling from altitude $h$". It satisfies: $r_i^t(d,\ h) = n_i^t(1 - S_i)P_{fatal}(d,h)P_{failure}$ where $n_i^t$ is the number of people at location $i$ at time $t$, $S_i$ is the percentage of area sheltered by the building at location $i$, $P_{fatal}(d,h)$ is the probability of fatality that a crash of UA weighing $d$ falling from altitude $h$ is fatal for an unprotected average person, and $P_{failure}$ is the probability that UA loses control with an uncontrolled descent with a crash on the ground.

# 4. Scenario analysis

$Population\ risk$ can be assessed based on various scenarios of UAS operations. Appropriate settings are required to create each scenario, such as the UA specifications, mission altitude, failure probability, and fatality probability. This section demonstrates a methodology for assessing $population\ risk$ using hypothetical scenarios of UAS operations in Seoul. We selected the Jung and Mapo districts for our case studies. Jung district is a major commercial area with a densely built environment, and Mapo district is one of the major residential areas, as shown in Fig. 2.

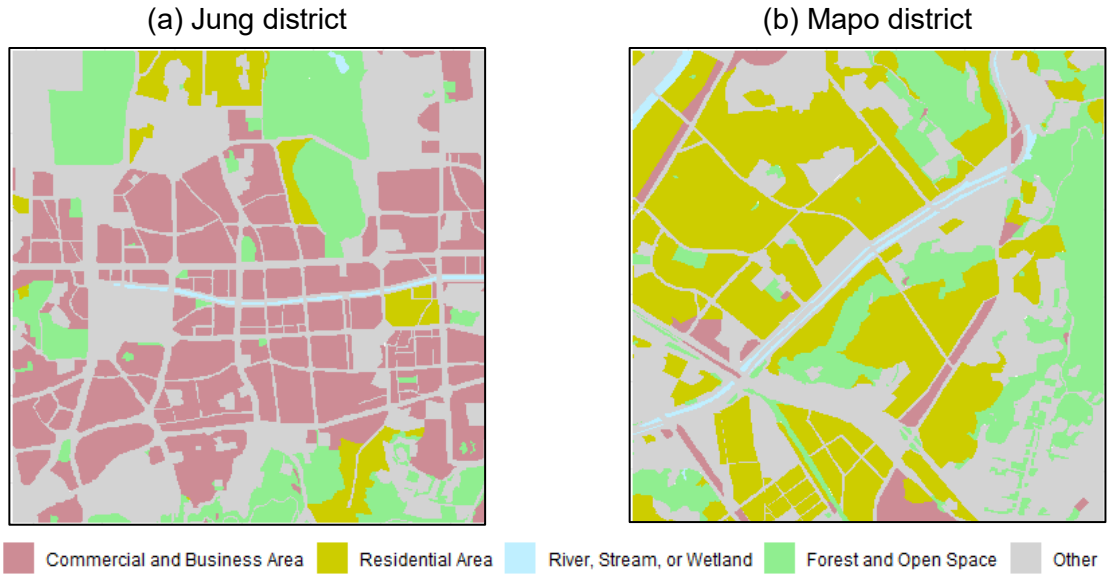

**Fig. 2.** Distribution of land use. (**a**) Jung district; (**b**) Mapo district

### 4.1 Scenario generation

The hypothetical scenarios are based on a multicopter UA weighing 1.1kg, operating at a height of 20m. A detailed description of the scenario settings is provided below.

We assumed that there would occur 30.23 system failures per million flight hours over a location $y$, in accordance with Petritoli et al. (2018). Using actual data, they reported the system failure rate of the UAS operations per flight hour. The probability of fatality was calculated using the KE. When the UA collides with a person on the ground, the KE is transferred from the UA to the person. Although the impact energy of a collision depends on many factors, several studies have attempted to estimate it. In this study, we adopted the findings of a notable study by Koh et al. (2018). Using UA-free drop modeling, they calculated the impact energy on a dummy head when falling from various heights. The energy was then converted into AIS scores to classify the injury levels. The fatality rate for a UA weighing 1.1 kg when falling from 20m was 8-10%. As a measure of the sheltering effect, we calculated the percentage of sheltered area within location $y$ using building data. Lastly, we used de facto population data to calculate the population density of location $y$ at time $t$.

Using these calculated risk values, we classified each district's airspace to identify restricted airspaces. Each district's airspace is regarded as "restricted" when the risk exceeds a threshold value of $10^{-7}$ per flight hour. This is the airworthiness certification standard for UA to prevent catastrophic accidents (King et al., 2005).

### 4.2 Empirical results and discussions

In this section, we present population risk maps and restricted airspace maps to explore the risk trends over time and geographical variations. When $population\ risk$ exceeds an acceptable level, airspace is restricted. This study adopted an acceptable level of safety of $10^{-7}$.

Figure 3 presents the empirical results of a case study conducted in Jung district, a commercial area. When the risk is derived from the residential population, high-risk areas are scattered across multiple areas (Fig. 3(a)). Consequently, the corresponding airspaces are restricted, occupying 2.58% of the region of interest, as shown in Fig. 3(d). High-risk areas can vary by time of day if the risk is derived from the de facto population data. From 03:00 to 04:00, high-risk areas were scattered across several areas, resulting in restricted airspace (Fig. 3(b)). The restricted airspaces are similar to the results derived from the residential population data. This may be because most people spend their nights in residential areas. In contrast, from 13:00 to 14:00, noticeable high-risk areas and restricted airspaces were observed, as shown in Fig. 3(c) and Fig. 3(f). It is apparent that the restricted airspaces are clustered around the center of the map compared to the other restricted maps. This may be due to the characteristics of the region, which is a major commercial area. In commercial areas, the population density tends to be higher during the day than at night. Moreover, the restricted airspace accounted for 2.32% from 03:00 to 04:00 and 36.07% from 13:00 to 14:00. In comparison with the results of restricted airspace based on residential population, the restricted airspace results from 03:00 to 04:00 and 13:00 to 14:00 show a difference of -0.26% and 33.50%, respectively. Over time, there is a change between restricting airspace based on the risk derived from the residential population and restricting airspace based on the risk derived from the de facto population.

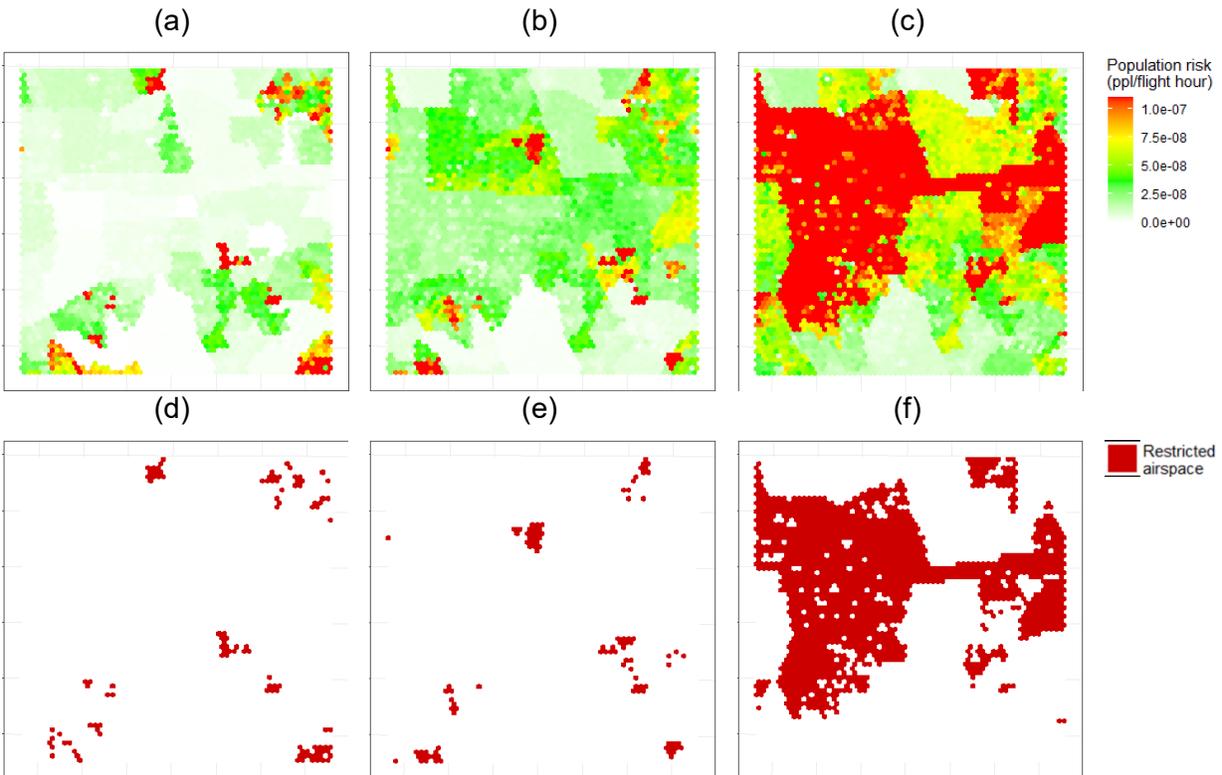

**Fig. 3.** Distribution of population risk and restricted airspace in the Jung district. (a) Residential population-derived population risk map; (b) de facto population-derived population risk map (03:00-04:00); (c) de facto population-derived risk map (13:00-14:00); (d) residential population-derived restricted airspace map; (e) de facto population-derived restricted airspace map (03:00-04:00); (f) de facto population-derived restricted airspace map (13:00-14:00).

Figure 4 presents the empirical results from a case study conducted in Mapo district, one of the residential areas. When the risk was derived from the residential population, high-risk areas were observed mainly in the residential areas (Fig. 4(a)). Accordingly, 14.15% of the airspaces are restricted. High-risk areas were also observed in residential areas when the risk was derived from de facto population data from 03:00 to 04:00 (Fig. 4(b)). During the night, people are mostly concentrated in residential areas. The spatial distribution of high-risk areas was similar regardless of time. Figure 4(c) shows the high-risk areas observed in residential areas when the risk was derived from the de facto population from 13:00 to 14:00. Rather, there was a slight decrease in high-risk areas compared with the results from 03:00 to 04:00. This seems to be because people move to work during the day. When the risk was derived from the de facto population from 03:00 to 04:00 and 13:00 to 14:00, the restricted airspace accounted for 15.38% and 12.33%, respectively. Compared with the results of the restricted airspace based on the residential population, the results of the restricted airspace from 03:00 to 04:00 and 13:00 to 14:00 show a difference of 1.23% and -1.81%, respectively. Some people may move to commercial work areas.

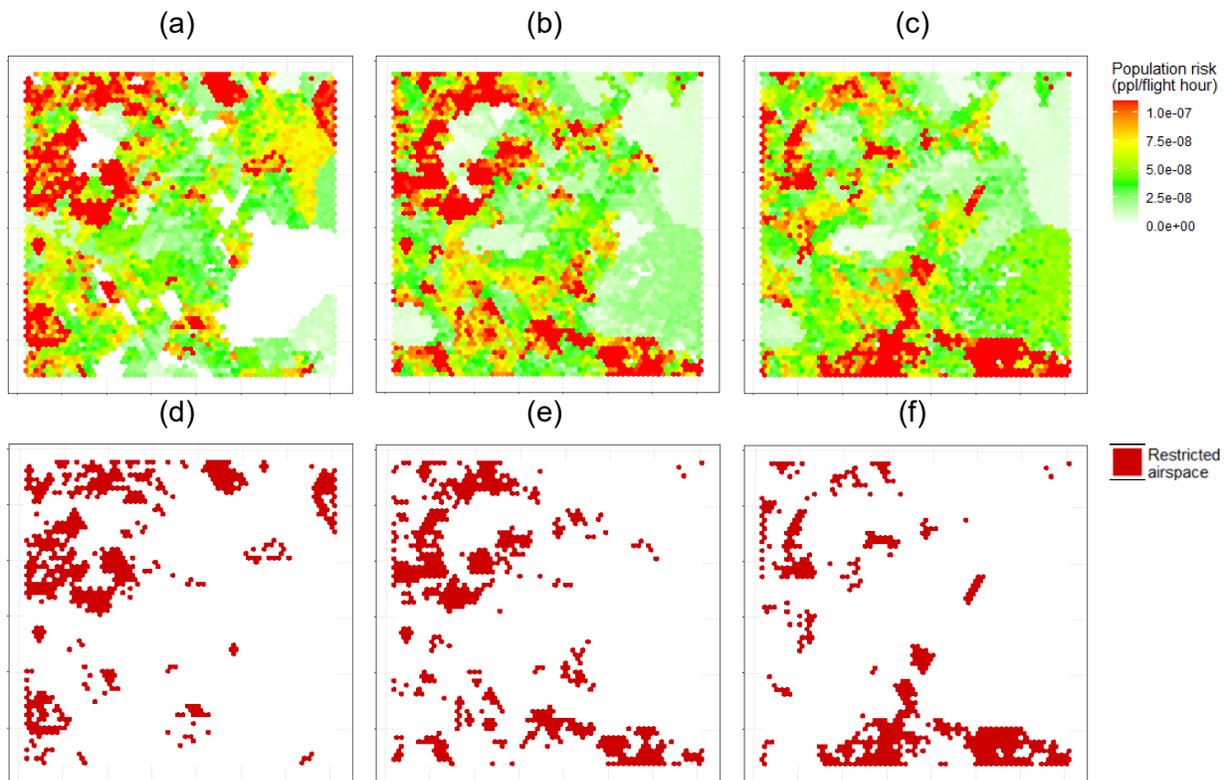

**Fig. 4.** Distribution of population risk and restricted airspace in the Mapo district. (**a**) Residential population-derived population risk map; (**b**) de facto population-derived population risk map (03:00-04:00); (**c**) de facto population-derived risk map (13:00-14:00); (**d**) residential population-derived restricted airspace map; (**e**) de facto population-derived restricted airspace map (03:00-04:00); (**f**) de facto population-derived restricted airspace map (13:00-14:00).

Figure 5 shows the percentage of restricted airspace observed during the day in the Jung and Mapo districts. The percentage was calculated as the ratio of the restricted airspace to the entire airspace.

In the Jung district, the restricted airspace ratio is highest from 14:00 to 15:00 and lowest from 7:00 to 8:00. The highest percentage was 36.61%, whereas the lowest was 2.02%. These results may be related to the characteristics of commercial districts, where the population density is higher during the daytime than at night. The restricted airspace ratio changed with the time of day. In the daytime, the percentage of restricted airspace tends to increase, whereas at night, it tends to decrease. The results suggest that it is necessary to consider the time of day when restricting airspace based on $population\ risk$. In comparison with restricting airspace based on the risk derived from the residential population, there was a significant difference during the daytime and a small difference during the nighttime. Restricting airspace based on the risk derived from the residential population may distort the effect of population density.

In Mapo district, the restricted airspace ratio is the highest from 0:00 to 1:00 (16.61%) and the lowest from 10:00 to 11:00 (12.24%). Compared to the results of Jung district, the difference over time is relatively small. There are changes in the restricted airspace with the time of day, but the ratio is higher at night than during the daytime. In the daytime, the percentage of restricted airspace tends to decrease, whereas at night, it tends to increase. Some people may move from residential areas to commercial areas during the daytime. Compared to restricting airspace based on the risk derived from the residential population, there is not much difference at any time of the day. The percentage of restricted airspace does not fluctuate abruptly but rather gradually over time. Similar values were observed at similar time intervals. This suggests that a reasonable period can be helpful when issuing restrictions on airspace.

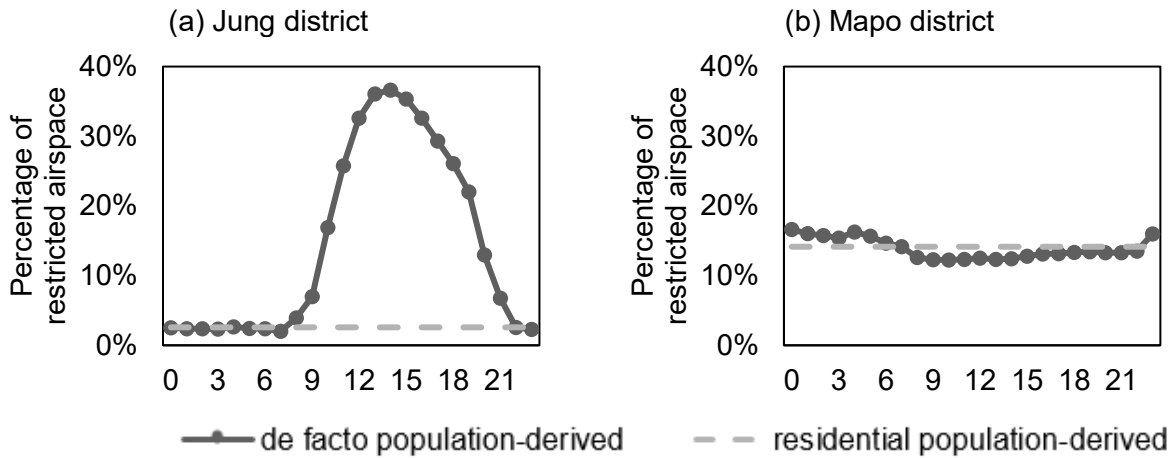

**Fig. 5.** Hourly restricted airspace profile using residential and de facto population data. (**a**) Jung district; (**b**) Mapo district

## 5. Sensitivity analysis

As mentioned in Section 2, $population\ risk$ is estimated using information gathered from existing datasets, published literature, and controlled experiments. In this approach, the input parameters are estimated with varying degrees of accuracy depending on the quality and availability of the data. However, there may be a lack of reliable empirical data for the input parameter values. In the presence of large degrees of uncertainty regarding input parameter estimates, inaccurate conclusions can be drawn. To interpret the importance of input parameters in determining outcome variability, it is critical to measure how variation propagates from the input parameters to the outcomes. As one of the methods to analyze how a variable's input affects an outcome, sensitivity analysis provides information about the relative importance of each variable in determining outcome variability for combinations of multiple input parameters (Borgonovo and Plischke, 2016; Frey and Patil, 2002; Morris, 1991). Consequently, it provides insight into the input variables that can influence the outcomes and which input variables require precise estimation for accurate outcomes. To identify input variables requiring precise estimation, we conducted sensitivity analysis using the Morris method and Risk Achievement Worth.

### 5.1 Morris method

This study utilized the Morris method, a widely used technique for sensitivity analysis, which provides the effect of changes in input parameter values on outcomes (Morris, 1991). It can capture the outcome variation due to the movement of a sampled point to an adjacent point. To sample a set of points, the input parameter was discretized by transforming it into a $p$ level interval variable. Based on the discretized input parameters, the magnitude of variation in the model output is calculated as follows until all possible input parameter values have been explored: (1) sample a set of initial values from a range of possible values for all inputs and calculate the corresponding output of the model; (2) change the value of one variable (all other features remain at their initial value) and calculate the changed output of the model compared to the first run; and (3) change the value of another variable (the previous feature remains at the changed value and all other features remain at their initial value) and calculate the changed output of the model compared to the second run.

This variation due to the defined variation of one input parameter $X$ is the elementary effect (EE): $EE_i = (Y(X + e_i\Delta_i) - Y(X))/\Delta_i$ where $e_i$ is a vector of zeros, except for the $i$-th input parameter that equals $\pm 1$. To evaluate the elementary effects for many combinations $n$, a summary statistic on the simulated elementary effects can be calculated and interpreted as a sensitivity indicator. The mean of the elementary effects $\mu_i$ is defined as $\mu_i = 1/n \sum_{j=1}^{k} EE_i^j$. This is interpreted as the average effect of the input variable $i$ on the model output $j$ variation. A high value of $\mu_i$ implies that input parameter $i$ is an influential variable on the model output, and a low value of $\mu_i$ implies that input parameter $i$ is a non-influential variable in the model output. The average absolute value of the elementary effects was defined as $\mu_i^* = 1/n \sum_{j=1}^{k} |EE_i^j|$. Compared with the mean $\mu_i$, the absolute mean $\mu_i^*$ is a more robust sensitivity indicator that ensures robustness against non-monotonic models. Similar to $\mu_i$, a high value of $\mu_i^*$ implies that input parameter $i$ is an influential variable on the model output, and a low value of $\mu_i^*$ implies that input parameter $i$ is a non-influential variable in the model output.

### 5.2 Risk Achievement Worth method

To measure the importance of a feature in achieving the present risk, we adopted one of the PSA importance measures, Risk Achievement Worth (RAW). The RAW is defined as the increase in risk if the feature is assumed to fail at all times. Let $R_i^*$ be the increased risk level with feature $i$ assumed to fail, and $R_0$ is the present risk level. Thus on a ratio scale, RAW is $A_i =$ $R_i^*/R_0$. RAW is calculated as follows: (1) sample a set of values within the possible values for all input variables and calculate the present risk level $R_0$, (2) remove one feature (all other features remaining at their values) and calculate the increased risk level with feature $i$ assumed to fail ($R_i^*$), and (3) calculate RAW $A_i = R_i^*/R_0$. These steps are repeated until all RAW values of the input variables are obtained. For many combinations $n$, a statistical measure for the evaluation of RAW can be calculated and interpreted as sensitivity: coefficient of variation $CV_i = \sigma_i/\mu_i$, where $\mu_i = (1/n)\sum_{j=1}^{k} A_i^j$ and $\sigma_i = \sqrt{(1/n-1)\sum_{j=1}^{k}(A_i^j - \mu_i)^2}$. A high value of $CV_i$ implies that input parameter $i$ is an important variable in the model output, and a low value of $CV_i$ implies that input parameter $i$ is an unimportant variable in the model output.

### 5.3 Results and discussions

Using the Morris and RAW methods, sensitivity analyses were conducted to identify the input variables that influenced $population\ risk$. Table 2 lists the input variables and parameter values used in this study. All possible parameter values were chosen within reasonable ranges based on data from existing datasets, published literature, and controlled experiments.

**Table 2.** Estimates of input parameters for sensitivity analysis

| Parameter | Data | Source |
| --- | --- | --- |
| Probability of accident (%) | $10^{-6}$ per flight hour (FH); $10^{-5}$ per FH; $10^{-4}$ per FH; $3.42 \times 10^{-4}$ per FH; $30.23 \times 10^{-6}$ per FH; | Clothier et al. (2007); Ford and McEntee (2010); Melnyk et al. (2014); Stevenson et al. (2015); Petritoli et al. (2018) |
| Probability of fatality (%) | 0.01; 0.02; 0.05; 0.08; 0.1; 0.5; 1 | Clothier et al. (2007); Lum et al. (2011); Burke et al. (2011); Melnyk et al. (2014); Koh et al. (2018) |
| Sheltering effect (%) | Building data | National Spatial Data Infrastructure Portal (2020) |
| Population density ($ppl/m^2$) | De facto population data | Seoul Open Data Plaza (2020) |

The results of the sensitivity analysis, including the Morris method and RAW, are shown in Fig. 6. In the Morris method, a high $\mu_i^*$ value indicates that the input variable has a greater effect on the model outcome. Population density and the sheltering effect had the greatest impact on $population\ risk$. The coefficient of variation ($CV_i$) in the RAW results indicates the effect of the input variable on the outcome. A high value of $CV_i$ implies that input parameter $i$ is an important variable in the model outcome. The probability of an accident and population density have the greatest impact on $population\ risk$. Both sensitivity analysis methods indicated that population density has a significant effect on the population risk associated with UAS operations. Overall, the ranking in Fig. 6, is consistent with findings from previous studies which identify UAS weight, population density, and vehicle failure rate as influential input variables for the risk of UAS operations (Melnyk et al., 2014).

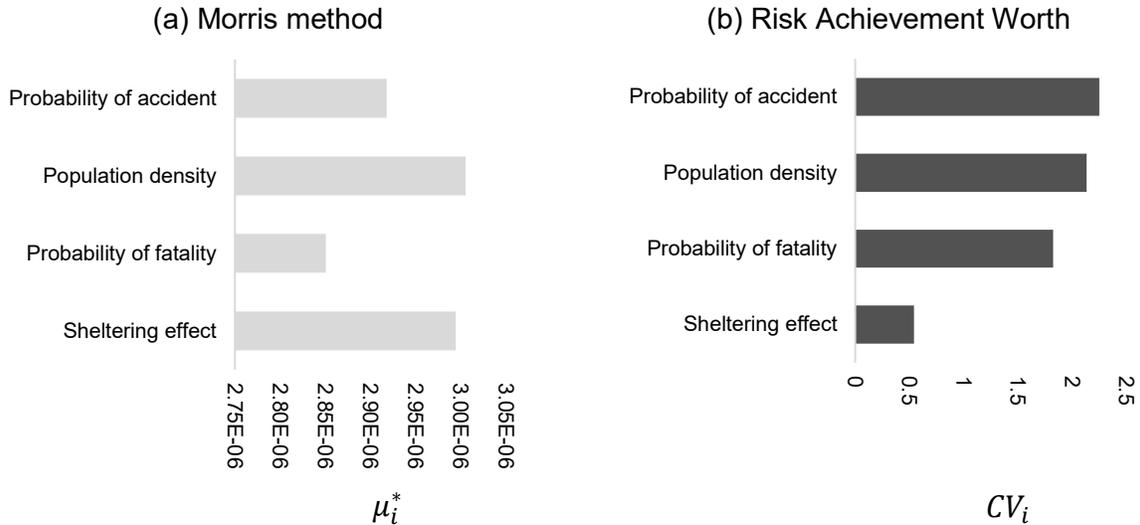

**Fig. 6.** The results of sensitivity analysis. (**a**) Morris method; (**b**) Risk achievement worth

## 6. Conclusion

In this study, we defined and measured a generic metric of population risk that enables us to incorporate spatiotemporal changes in the number of people who are actually exposed to UAS operations. By utilizing high-resolution de facto population data, we calculated the population risk of UAS operations at each time of day. Compared with the risk values derived from residential population data, our metric reflected spatiotemporal characteristics of population distribution when assessing the population risk associated with UAS operations.

Moreover, we investigated changes in airspace operability throughout the day in terms of shape and volume by restricting the airspace areas in which the risk value exceeded the threshold. Based on the scenario analysis in Seoul, including the Jung and Mapo districts, we found that there is a difference in the restricted airspace ratios by the time of day. Especially during the daytime, restricted airspaces were clustered around the business areas. We also found a difference between restricting airspace based on residential population-derived risk and de facto population-derived risk. This is evidence of the importance of accurately estimating population density by time of day when assessing the population risk of UAS operations. Our sensitivity analysis also revealed the importance of accurately estimating the population density for multiple combinations of input parameter values.

One limitation of our approach is that it does not consider the traffic density of the UAS operations. Without considering the traffic density of UAS operations, we conservatively restrict the airspace as a whole. Future studies may benefit from the realistic traffic density of UAS operations to define more reasonably restricted airspace boundaries.

## CRediT authorship contribution statement

The authors confirm contribution to the paper as follows: study design: S. Oh and Y. Yoon; data collection: S. Oh; analysis and interpretation of results: S. Oh and Y. Yoon; draft manuscript preparation: S. Oh. All authors reviewed the results and approved the final version of the manuscript

## Acknowledgements

"This work is financially supported by Korea Ministry of Land, Infrastructure and Transport (MOLIT) as 「Innovative Talent Education Program for Smart City」".